\documentclass{article-hermes}

\input boxedeps.tex
\input epsf.tex
\SetRokickiEPSFSpecial
\HideDisplacementBoxes
\def\drv#1{\buildrel#1\over \longrightarrow}

\def\QED{\hbox{\rlap{$\sqcap$}$\sqcup$}}
\def\sl{\mathrel{< \!\!\!< }}
\def\sr{\mathrel{> \!\!\!> }}

\def\qed{\ifmmode\eqno\QED\else{\parfillskip=0pt \enspace\hfill
  \rlap{$\sqcap$}$\sqcup$ \ifdim\lastskip<\medskipamount
  \removelastskip\penalty55\medskip\fi \parfillskip=0pt
  plus1fil}\fi}

\newenvironment{mitemize}{\begin{itemize}\setlength{\itemsep}{0pt}
\setlength{\parsep}{0pt}
\setlength{\partopsep}{-16pt}
\setlength{\topsep}{-15pt}
\setlength{\parskip}{0pt}
\setlength{\labelwidth}{1pt}}
{\end{itemize}}

\newenvironment{menumerate}{\begin{enumerate}\setlength{\itemsep}{0pt}
\setlength{\parsep}{0pt}
\setlength{\partopsep}{-15pt}
\setlength{\topsep}{-15pt}
\setlength{\parskip}{0pt}
\setlength{\labelwidth}{1pt}}
{\end{enumerate}}

\newtheorem{definition}{Definition}
\newtheorem{thm}{Theorem}

\newtheorem{ex}{Example}

\def\proof{\medskip\noindent {\bf Proof: }}

\def\A{{\cal A}}

{
}
\frenchlangfalse

\begin{document}

\journal{CSIT'05}{}{}{}{-}{-}
\frenchlangfalse
\title[Multiple serial episodes matching]
{Multiple serial episodes matching\fup{****}}
  
\date{\today}
\vskip -1cm

\author{Patrick C\'egielski \fup{*} \andauthor Ir\`ene~Guessarian \fup{**}
\andauthor Yuri~Matiyasevich \fup{***}}

\address{
\fup{*}LACL, UMR-FRE 2673, Universit\'e~Paris~12, Route
fores\-ti\`ere Hurtault, F-77300 Fontainebleau, France,\\
cegielski@univ-paris12.fr \\[3pt]
\fup{**}LIAFA, UMR 7089 and~Universit\'e~Paris~6, 2 Place
Jussieu, 75254 Paris Cedex 5, France; 
send correspondence  to
ig@liafa.jussieu..fr\\[3pt]
\fup{***} Steklov Institute of Mathematics,
Fontanka 27, St. Petersburg, Russia.
yumat@pdmi.ras.ru\\[3pt]
\fup{****} Support by INTAS grant 04-77-7173 is gratefully acknowledged.}


\abstract{In \cite{bcgm} we have generalized the Knuth-Morris-Pratt (KMP)
pattern matching algorithm and defined a non-conventional kind of RAM, the
MP--RAMs which model more closely the microprocessor operations, and designed
an $O(n)$ on-line algorithm for solving the serial episode matching problem on
MP--RAMs when there is only one single episode. We here give two extensions of
this algorithm to the case when we search for several patterns simultaneously
and compare them. More preciseley, given $q+1$ strings (a text $t$ of length
$n$ and $q$ patterns $m_1,\ldots,m_q$) and a natural number $w$, the
{\em multiple serial episode matching problem} consists in finding the number of size
$w$ windows of text $t$ which contain patterns $m_1,\ldots,m_q$ as
subsequences, i.e. for each $m_i$, if $m_i=p_1,\ldots ,p_k$, the letters
$p_1,\ldots ,p_k$ occur in the window, in the same order as in $m_i$, but not
necessarily consecutively (they may be interleaved with other letters).}

 
\maketitlepage

\section{Introduction} 

The recent development of datamining induced the development of computing
techniques, among them is episode searching and counting. An example of
frequent serial episode search is as follows: let $t$ be a text consisting of
requests to a  university webserver ; assume we wish to count how many times,
within at most 10 time units, the  sequence $e_1e_2e_3e_4$ appears, where
$e_1=$ `Computer Science', $e_2=$ `Master', $e_3=$ `CS318 homepage',
$e_4=$ `Assignment'. It suffices to count the number of 10-windows of $t$ 
containing the subsequence $p=e_1e_2e_3e_4$. If $e_1,e_2,e_3,e_4$ must appear
in that same order in the window, the episode is said to be {\em serial}, if
they can appear in any order, the episode is said to be {\it parallel}; a
partial order can also be imposed on the events composing an episode (see
\cite{mtv}, which proposes several algorithms for episode searching).
Searching serial episodes is more complex than searching parallel episodes. Of
course, if one has to scan a $\log$ file, it is better to do it for several
episodes $e_1e_2\ldots e_n$, $f_1f_2\ldots f_m$, $g_1g_2\ldots g_p$
simultaneously. We will hence investigate the search of several serial episodes
in the same window: each serial episode is ordered, but no order is imposed
among occurrences of the episodes in the window.

 
The  problem we address is the following: given a text $t$  of length $n$,
patterns $m_1,\ldots, m_q$ on the same alphabet $A$ and an integer $w$, we
wish to determine the number of size $w$ windows of text containing all $q$
patterns as serial episodes, {\it i.e.} the letters of each $m_i$ appear in the
window, in the same order as in $m_i$, but they need not be consecutive
because other letters can be interleaved. When searching for a single pattern
$m$, this  problem with arguments the window size $w$, the text $t$ and
pattern $m$ is called {\em serial episode matching problem } in \cite{mtv},
{\em  episode matching} in \cite{dfggk} and {\em  subsequence matching} in \cite{ahu};
a related problem is the  {\em  matching with don't cares} of \cite{mby,kr}. 

This problem is an interesting generalisation of {\em pattern-matching}. Without
the window size restriction, it is easy to find in linear time whether $p$ 
occurs in the text: if $p=p_1\ldots p_k$, a finite state automaton with $k+1$
states $s_0,s_1,\ldots,s_k$ will read the text; the initial state
is $s_0$; after
reading letter $p_1$ we go to state $s_1$, then after reading letter $p_2$ we
go to state $s_2$, \dots; the text is accepted as soon as state $s_k$ is
reached. Episode matching within a $w$-window is harder; its importance is
due to potential applications to datamining \cite{m,mtv} and molecular
biology\cite{mby,kr,nr}.

For the  problem with a single episode in $w$-windows, a standard algorithm
is described in \cite{dfggk,mtv}. It is close to the algorithms of
{\em pattern-matching} \cite{a,ahu} and its time complexity is $O(nk)$. Another
{\em on-line} algorithm is described in \cite{dfggk}: the idea is to slice the
pattern in $k/\log k$ well-chosen pieces organised in a {\em trie}; its time
complexity is $O(nk/\log k)$. We gave an {\em on-line} algorithm reading the
text $t$, each text symbol being read only once and whose time complexity  
is $O(n)$ \cite{bcgm}.

In this paper, we describe two efficient algorithms (Section
\ref{sect.plusmot}) for solving the problems of simultaneous search of multiple
episodes. These algorithms use the {\em MP--RAM}, that we introduced in
\cite{bcgm}, to model microprocessor basic operations, using only the fast
operations on bits ({\em shifts}), and bit-wise addition; this gives an
on-line algorithm in time $O(nq)$ (theorem \ref{mult.thm}). In practice, this
algorithm based on MP--RAMs and a new implementation of {\em tries}, is much
faster as shown in section \ref{sect.exp}. We believe that other algorithms
can be considerably improved if programmed on MP--RAMs.

Our algorithm relies upon two ideas: 1) preprocess patterns and window size
to obtain a finite automaton solving the problem as in Knuth, Morris, and
Pratt algorithm \cite{kmp} (the solutions preprocessing the text
\cite{t,mby,sliss,u} are prohibitive here because of their space complexity)
and 2) code the states of this automaton to compute its transitions very
quickly on MP-RAMs, without precomputing, nor storing the automaton: using
the automaton itself 
is also prohibitive, not the least because of the number of
states; we emulate the behaviour of the automaton without computing the
automaton. We  study: (a) the  case when the patterns have no common part
and (b) the case when they have similar parts. In each case, an appropriate
preprocessing of the set of patterns enables us to build an automaton solving
the problem and we show that the behaviour of this automaton can be emulated
on-line on MP-RAMs. Moreover, the time complexity of the preprocessing is
insignificant because it is smaller than the text size by several orders of
magnitude: typically, window and patterns will consist of a few dozen
characters while the text will consist of several million characters.

The paper is organised as follows: in section 2, we define the problem, in
section \ref{sect.plusmot} we describe the algorithms searching multiple
episodes in parallel; we present the experimental results in section
\ref{sect.exp}.

\section{The problem}

\subsection{The  (multiple) episode problem}

An {\it alphabet} is a finite  non-empty set $A$. A {\it length} $n$ {\it word}
on $A$ is a mapping $t$ from $\{1,\ldots,n\}$ to $A$. The only length zero
word is the {\it empty word}, denoted by $\varepsilon$. A non-empty word \quad
$t \; : \; i \mapsto t_i$\quad is denoted by $t_1t_2 \cdots t_n$. A
{\it language} on alphabet $A$ is a set of words on $A$.

Let $t = t_1t_2 \cdots t_n$ be a word which will be called the {\em text} in
the paper. The word $p = p_1p_2 \cdots p_k$ is a {\it factor} of $t$ iff, there 
exists an integer $j$ such that $t_{j+i} = p_i$ for $1 \leq i \leq k$. A size
$w$ {\it window} of on $t$, in short $w$-window, is a size $w$ factor
$t_{i+1}t_{i+2} \cdots t_{i+w}$ of $t$; there are $n-w+1$ such windows in
$t$. The word $p$ is an {\it  episode} (or {\it subsequence}) of $t$ iff there
exist integers $1 \leq i_1 < i_2 < \cdots < i_k \leq n$ such that
$t_{i_j} = p_j$ for $1 \leq j \leq k$. If moreover, $i_k - i_1 < w$, $p$ is an
{\it  episode} of $t$ {\it in a $w$-window}. 

\begin{ex}\label{ex1} {\rm If $t={}$``dans ville il y a vie" (a French
advertisement, see figure \ref{monoprix.fig}) then ``vie" is a factor and hence a subsequence of $t$.
``vile" is neither a factor, nor a subsequence of $t$ in a $4$-window, but it
is a subsequence of $t$ in a $5$-window. See figure \ref{monop.fig}.}\QED
\end{ex}

\begin{figure}
\BoxedEPSF{monoprix.eps scaled 400}
\caption{\small A French advertisement}
\label{monoprix.fig}
\end{figure}

\begin{figure}
\BoxedEPSF{monop2.eps scaled 450}
\caption{\small A text with two $5$-windows containing ``vie" (in gray), and a
single $5$-window containing ``vile".}
\label{monop.fig}
\end{figure}

\smallskip

\noindent Given an alphabet $A$, and words $t,m_1,\ldots, m_q$ on $A$:

\begin{mitemize}

\item the simultaneous {\em pattern-matching} problem consists in finding whether
$m_1,\ldots,m_q$ are factors of $t$,

\item given moreover a window size $w$: 

\begin{mitemize}

\item the {\em subsequence existence} problem consists in finding whether
$m_1,\ldots,m_q$ are subsequences of $t$ in a $w$-window;

\item the {\em  multiple episode search} problem consists in  counting the number
of $w$-windows in which all of $m_1,\ldots,m_q$ are subsequences of $t$.

\end{mitemize}
\end{mitemize}

For the simultaneous search of several subsequences $m_1,\ldots,m_q$, we
have various  different problems:

\begin{itemize}

\item either we count the number of occurrences of each $m_i$ in a
$w$-window (not necessarily the  same): this case will be useful for
searching in parallel, with a single scan of the text, a set of patterns
which are candidates for being frequent.

\item or we count the number of windows containing all the $m_i$s: this case
will be useful for trying to verify association rules. For example, the
association rule $m_2,\ldots,m_q \Longrightarrow m_1$ will be useful if the
number of $w$-windows containing all the $ m_2,\ldots,m_q$ is high enough, and
to check that, we will count the $w$-windows containing {\em all of}
$m_2,\ldots,m_q$. Our method will enable us to verify more easily both the
validity of the association rule (``among the windows containing
$ m_2,\ldots,m_q$ many contain also $m_1$'') and the fact that it is
interesting enough (``many windows contain $m_2,\ldots,m_q$''): it will
suffice to count simultaneously the windows containing $m_2,\ldots,m_q$ and
the windows containing $m_1, m_2,\ldots,m_q$.

\end{itemize}

A naive solution exists for {\em pattern-matching}. Its time complexity on RAM is
$O(nk)$, where $k$ is the pattern size. Knuth, Morris, and Pratt
\cite{kmp} gave a well-known algorithm solving the problem in linear time
$O(n + k)$. A solution in $O(nk)$ is given in \cite{mtv} for searching a single
size $k$ episode. We gave in \cite{bcgm} an algorithm with time complexity
$O(n)$ (on MP--RAM) for searching a single episode.

\subsection{The notation  $o(nk)$}

Let us first make precise the meaning of the notation $o(nk)$.

The notation $o(h(n))$ was introduced to compare growth rates of functions
with one argument; for comparing functions with several arguments, various
non-e\-qui\-va\-lent interpretations $o(h(n,m,...))$ are possible. Consider a
function $t(n,k)$;  $t(n,k)=o(nk)$ could mean:

\begin{menumerate}

\item either $\displaystyle \lim_{n+k\to+\infty} t(n,k)/nk =0$;

\item or
$\displaystyle\lim_{n\to+\infty\atop k\to+\infty} t(n,k)/nk =0$, i.e. 
$\forall \epsilon, \exists N, \ \forall n,\forall k\allowbreak
\big(n>N \hbox{ and  }k>N\allowbreak \Longrightarrow t(n,k)<\epsilon nk\big)$.

\end{menumerate}
\vspace{ -5pt}

With meaning 1, no algorithm can solve the {\it single
episode within a window} problem
in time $o(nk)$. Indeed, any algorithm for the {\it episode within a  window}
problem must scan the text at least once, hence $t(n,k)\geq n$. For a given
$k$, for example $k=2$,we have $t(n,k)/nk\geq 1/2$. Hence
$\displaystyle \lim_{n+k\to+\infty} t(n,k)/nk =0$ is impossible. We thus have
to choose meaning 2.

\subsection{Algorithms on MP--RAM}
\label{section.algo3}

Given a window size $w$ and $q$ patterns, we preprocess (patterns + window
size $w$) to build a virtual finite state automaton $\A$; we will then emulate
on-line the behaviour of $\A$ to scan text $t$ and count in time $nq$ the
number of windows containing our patterns as episodes. Note that our method is
different from both: 1) methods preprocessing the text \cite{t,mby,sliss,u} (we
preprocess the pattern) and 2) methods using suffixes of the pattern
\cite{croch,mby,kr,u} (we use prefixes of the patterns). We encode the subset
of states of $\A$ needed to compute the transitions on-line on an MP-RAM.
Indeed, $\A$ has $O(w+1)^k$ state, where $k$ is the size of the structure
encoding the $q$ patterns $m_1,\ldots,m_q$; for $w$ and $q$ large, the time
and space complexity for computing the states of $\A$ becomes prohibitive,
whence the need to compute the states on-line quickly without having to
precompute nor store them. We introduced MP-RAMs to this end.

{\em Pattern-matching} algorithms are often given on RAMs. This model is not good
when there are too many different values to be stored, for example
${O(w+1)^ k}$ states for $\A$. As early as 1974, the motivation of \cite{prs}
for introducing ``vector machines'' was the remark that boolean bit-wise
operations and shifts which are implemented on computers are faster and better
suited for many problems. This work was the starting point of a series of
papers: \cite{trl,bg} comparing the complexities of computations on various
models of machines allowing for boolean bit-wise operations and shifts with
computation complexities on classical machines, such as Turing machines, RAMs
etc. The practical applications  of  this technique to various
{\em pattern-matching} problems start with \cite{byg,wm}: they are known as
{\it bit-parallelism}, or {\it shift-OR} techniques. We follow this track with the
episode search problem, close to the problems studied in \cite{byg,wm,byn},
albeit different from these problems.

In the sequel, we use a variant of RAMs, which is a more realistic computation
model in some aspects, and we encode $\A$ to ensure that (i) each state of
$\A$ is stored in a single memory cell and (ii) only the most basic
microprocessor operations are used to compute the transitions of $\A$. Our
RAMs have the same control structures as classical RAMs\footnote{See
\cite{ahu} pages 5--11, for a definition of classical RAMs.}, but the
operations are enriched by allowing for boolean bit-wise operations and
shifts, which we will preferably use whenever possible. Such RAMs are close to
microprocessors, this is why we called them MP--RAMs. 

\begin{definition} An MP--RAM is a RAM extended by allowing new operations:

\begin{menumerate}
\item the bit-wise {\rm and}, denoted by $\&$,
\item the {\em left shift}, denoted by $\sl$ or $shl$, and  
\item the {\em right  shift}, denoted by $\sr$ or $shr$.
\end{menumerate}
\end{definition}

The new operations are low-level operations, executable much faster than the
more complex {\it MULT, DIV} operations.

\begin{ex} 
Assume our ~\hbox{MP--RAMs}~have unbounded memory cells. We
will have for~example: $(10110 \ \& \ 01101) = 100$,
$(10110 \sl 4) =101100000$ and $(10110\sr 3) = 10$. If  memory cells have at most 8
bits, we will have: $(10110 \sl 4) = 1100000$, that will be written as
$(00010110 \sl  4) =01100000$.
\end{ex}

\section {Parallel search of  several patterns}
\label{sect.plusmot}

Let us recall the problems. Given patterns $m_1, m_2,\ldots,m_q$, we can:

\begin{itemize}

\item either count the number of occurrences of each $m_i$ in a $w$-window
(not necessarily the same one);

\item or count the number of $w$-windows containing $m_1,m_2,\ldots,m_q$. 

\end{itemize}  

The algorithm we described in \cite{bcgm} for counting the number of
$w$-windows containing {\em a single} pattern $m$ can be adapted to all these
cases, only the acceptance or counting condition will change.

To search simultaneously several patterns $m_1,\ldots,m_q$, \cite{wm} propose
a method concatenating all the patterns. To search simultaneously several
episodes $m_1,\ldots,m_q$, we generalise our algorithm \cite{bcgm}: we use $q$
counters $c_1,\ldots,c_q$ initially set to 0, and we define an appropriate
multiple counting condition such that each time $m_i$ is in a $w$-window, the
corresponding counter $c_i$ is incremented. This method has a drawback: if the
patterns are too long, it will need more than one memory cell for coding the
states of the automaton. For searching multiple patterns the method proposed
by \cite{dfggk} to optimise the search, when words $m_1,\ldots,m_q$ have
common prefixes, is to organise $m_1,\ldots,m_q$ in a {\em trie} \cite{k}
before applying the standard algorithm. We apply our algorithm on MP-RAMs in a
similar way, and implement {\em tries} in a new way. We thus can encode the set
of patterns compactly, and then encode the states of the automaton on a single
memory cell.

\subsection{Representing patterns by a trie}

Consider for example episodes $m_1=tu$, $m_2=tue$, and $m_3=tutu$. We choose
this example because it illustrates most of the difficulties in encoding the
automaton: episode {\it taie } is very simple because all letters are
different, {\it tati } is less simple because there are two occurrences of $t$
which must be distinguished, {\it tutu } a bit more complex (the first
occurrence of {\it tu} must be distinguished from the second one), 
{\it turlututu} would be even more complex. We represent these three episodes by
the trie $t$ pictured in figure \ref{trie.fig}.

\begin{figure}
\begin{center}
\BoxedEPSF{trie.eps scaled 460}
\caption{\small Trie representing $tu$, $tue$ and $tutu$. The full black
circles indicate ends of patterns.}
\label{trie.fig}
\end{center}
\end{figure}

We implement this trie $t$ by the three tables below:

\begin{center}
\begin{tabular}{r|c|c|c|c|c|} \cline{2-6}
\rule[0.4cm]{0cm}{2pt}$tr=$&t&u&e&t&u\\ \cline{2-6}
\end{tabular}\quad
\begin{tabular}{r|c|c|c|c|c|}\cline{2-6}
\rule[0.4cm]{0cm}{2pt}$pr=$&0&1&2&2&4\\ \cline{2-6}
\end{tabular}\quad
\begin{tabular}{r|c|c|c|}\cline{2-4}
\rule[0.4cm]{0cm}{2pt}$f=$&2&3&5\\ \cline{2-4}
\end{tabular}
\end{center}

Table $tr$ represents the ``flattened'' trie. Predecessors are in table $pr$:
$pr[i]$ gives the index in $tr$ of the parent of $tr[i]$ in the trie; 0 means
there is no predecessor and hence it is a pattern start\footnote{Numbering of
indices starts at 1 in order to indicate pattern starts by 0.}. Finally $f$
marks patterns ends: $f[i]$ is the index in $tr$ of the end of pattern $i$.

\subsection{Preprocessing the trie and algorithm}

We preprocess the trie of patterns and this gives us a finite state automaton
$\A$. Its alphabet is $A$. The states are the $k$-tuples of integers
$\langle l_1,\ldots,l_k\rangle$ with $l_j$ belonging to
$\{1,\ldots,w, +\infty\}$, where $k$ is the size of 
table $tr$  and $w$ the window
size. 

We describe informally the behaviour of $\A$. $\A$ scans $t$, it will be in 
state $\langle l_1,\ldots,l_k\rangle$ after scanning $t_1\ldots t_m$ iff
$l_i$ is the length of the shortest suffix\footnote{Word $s$ is a {\em prefix}
(resp. {\em suffix}) of word $t$ iff there exists a word $v$ such that $t=sv$
(resp. $t=vs$).} of $t_1\ldots t_m$ shorter than $w$ and containing
$tr[j_i]\ldots tr[i]$ as subsequence for $i= 1,\ldots,k$, where
$tr[j_i]$ $\ldots$ $tr[i]$ is the sequence of letters labelling the path going
from the root of the trie to the node represented by $tr[i]$. If no suffix (of
length less than $w$) of $t_1\ldots t_m$ contains $tr[j_i]\ldots tr[i]$ as a
subsequence, we let $l_i=+\infty$.

Let us now describe our algorithm. Let $\Omega$ be the least integer such that
$w+2\leq 2^\Omega$. The r\^ole of $+\infty$ is played by $2^\Omega-1$, whose
binary encoding is a sequence of $\Omega$ ones. We define the function
 Next$_{\Omega}$ by:
$${\rm Next}_{\Omega}(l)=\cases{l+1,&if $l<2^\Omega-1$;\cr
 2^\Omega-1,&else.\cr}$$

State $\langle l_1,\ldots,l_k\rangle$ is encoded by integer:
\begin{equation}
L=\sum_{i=1}^k l_i(2^{\Omega+1})^{i-1}
=\sum_{i=1}^k \Big( l_i\sl
\big((\Omega+1)(i-1)\big)\Big).\label{codstate.eq}
\end{equation}

\begin{figure*}
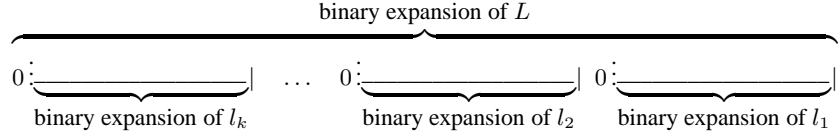

{\small
$$\overbrace{0\,\vdots\!\underbrace{\raise 2pt\hbox{\_\_\_\_\_\_\_\_\_\_\_\_\_\_\_\_\_\_}}_{\hbox{\footnotesize  binary expansion of $l_k$}}\!|\ \ \ \ \ldots\ \ \ \ 0\,\vdots\!\underbrace{\raise 2pt\hbox{\_\_\_\_\_\_\_\_\_\_\_\_\_\_\_\_\_\_}}_{\hbox{\footnotesize  binary expansion of $l_2$}}\! |\ \ 0\,\vdots\!\underbrace{\raise 2pt\hbox{\_\_\_\_\_\_\_\_\_\_\_\_\_\_\_\_\_\_}}_{\hbox{\footnotesize binary expansion of $l_1$}}\!|}^{\hbox{\footnotesize binary expansion of $L$}}$$
}
\caption{\small Encoding of  $\langle l_1,\ldots,l_k\rangle$.}
\label{es.fig}
\end{figure*}

Let $\overline{l_i}$ denote the binary expansion of $l_i$, $i=1,\ldots,k$,
 prefixed by zeros in such a way that $\overline{l_i}$
occupies $\Omega$ bits (all $l_i$s are smaller than $2^\Omega-1$,
 hence they will
fit in  $\Omega$ bits).
 The binary expansion of $L$ is obtained by concatenating the 
$\overline{l_i}$s, each prefixed by a zero
 (figure \ref{es.fig}).  These
initial zeros are needed for implementing function {Next}$_{\Omega}$ to
indicate overflows. Every integer smaller than $2^{k(\Omega+1)}$ can be written
as $k$ {\em big blocks} of $(\Omega+1)$ bits,
 the first bit of each big block is
0 (and is called the {\em overflow bit}) and the $\Omega$ remaining bits
constitute a {\em small block}. The blocks are numbered 1 to $k$ from right to
left (the rightmost block is block 1, the leftmost block is block $k$).

\begin{figure*}\label{ex.fig}
\begin{center}\begin{tabular}{r|r@{\small\ $\vdots$ }l|r@{\small\
  $\vdots$ }l|r@{\small\ $\vdots$ }l|r@{\small\ $\vdots$ }l|r@{\small\
 $\vdots$ }l|}\cline{2-11}
\rule[0.4cm]{0cm}{2pt}$L=$&0&$\overline {l_5}$&0&$\overline {l_4}$&0&$\overline
  {l_3}$&0&$\overline  {l_2}$&0&$\overline  {l_1}$\\ \cline{2-11}
\end{tabular}\end{center}
\caption{Encoding of  $\langle l_1,\ldots,l_5\rangle$; $\overline
  {l_i}$ is the binary expansion of $l_i$.}\label{fig0}
\end{figure*}

By the definition in equation~(\ref{codstate.eq}), the initial state
$\langle +\infty,\ldots,+\infty\rangle$ is encoded by:
$$ 
I_0=\sum_{i=1}^k
(2^\Omega-1)2^{(\Omega+1)(i-1)}
=\sum_{i=1}^k \Big(\big((1\sl \Omega )-1\big)\sl
{\big((\Omega+1)(i-1)\big)\Big)}.$$

One might see a multiplication here. In fact we will need a loop for $i=1$ to
$k$. We will execute each time we go through the loop a shift of $\Omega +1$,
and the multiplication will disappear. All equations below are treated in the
same way.

Assume that the window size is $w=13$ hence $\Omega = 4$. With the notations
of figure \ref{fig0}, state $l=\langle 2,5,\infty,5,\infty\rangle$ is encoded
by:
\begin{center}\begin{tabular}{r|r@{\small\ $\vdots$ }l|r@{\small\
        $\vdots$ }l|r@{\small\ $\vdots$ }l|r@{\small\ $\vdots$
        }l|r@{\small\ $\vdots$ }l|}
\cline{2-11}
\rule[0.4cm]{0cm}{2pt}$L=$&0&$\overline 
 {15}$&0&$\overline  {5}$&0&$\overline 
 {15}$&0&$\overline  {5}$&0&$\overline 
 {2}$\\ \cline{2-11}
\end{tabular}
\end{center}

The initial state is represented by:

\begin{center}\begin{tabular}{r|r@{\small\ $\vdots$ }l|r@{\small\
        $\vdots$ }l|r@{\small\ $\vdots$ }l|r@{\small\ $\vdots$
        }l|r@{\small\ $\vdots$ }l|}\cline{2-11}
$I_0=$&0&1111&0&1111&0&1111&0&1111&0&1111\\ \cline{2-11}
\end{tabular}\end{center}
or, writing $\underline {1}$ instead of the $\Omega$ ones representing
$\infty$:
\begin{center}\begin{tabular}{r|r@{\small\ $\vdots$ }l|r@{\small\
        $\vdots$ }l|r@{\small\ $\vdots$ }l|r@{\small\ $\vdots$
        }l|r@{\small\ $\vdots$ }l|}
\cline{2-11}
\rule[0.4cm]{0cm}{2pt}$I_0=$&0&$\underline  {1}$&0&$\underline 
 {1}$&0&$\underline  {1}$&0&$\underline  {1}$&0&$\underline 
 {1}$\\ \cline{2-11}
\end{tabular}
\end{center}

In transition $l=\langle l_1,\ldots,l_k\rangle \drv \sigma l'=\langle
 l'_1,\ldots,l'_k\rangle$, the $l'_i$ component of the new state $l'$ is either
{Next}$_\Omega(l_{pr[i]})$ or {Next}$_\Omega(l_{i})$ according to
whether the scanned letter $\sigma$ is equal to $tr[i]$ or not. The cases
$l'_i= {Next}_\Omega(l_{pr[i]})$ and $l'_i= {Next}_\Omega(l_{i})$
respectively yield a {\em first type computation} and a {\em  second type computation}.

To generalise the algorithm of \cite{bcgm}, we must define several {\em masks}
$M_\sigma$ for each letter $\sigma$ of alphabet A. If $\sigma$ has several
occurrences in table $tr$, we will need as many masks $M_\sigma$ as
occurrences $tr[i]$ and $tr[i']$ of $\sigma$ with $j=i-pr[i]\not=i'-pr[i']=j'$
(a single mask will suffice for the set of all occurrences such that $i-pr[i]$
has the same value $j$, because they correspond to the same shift of $j$ big
blocks). The $M_\sigma^j$ are the masks preparing first type computations.
Precisely, if $tr[i]=\sigma$ and $i-pr[i]=j$, the operation
$\big(L\sl j(\Omega+1)\big)\& M_\sigma^j$ will shift everything of $j$ big blocks
leftwards and will erase the blocks for which $\sigma\not= p_i$ or
$i-pr[i]\not=j$. For $i>1$, the $i$-th block will thus contain
$\overline{l_{pr[i]}}$ iff $tr[i]=\sigma$ and $i-pr[i]=j$. It will contain
\underline 0 otherwise.

In our example ($m_1=tu$, $m_2=tue$, and $m_3=tutu$), we will need two masks
$M_t$ but a single mask $M_u$ will suffice:

\begin{center}\begin{tabular}{r|r@{\small\ $\vdots$ }l|r@{\small\
 $\vdots$ }l|r@{\small\ $\vdots$ }l|r@{\small\ $\vdots$ }l|r@{\small\
 $\vdots$ }l|}\cline{2-11}
$M_t^1=$&0&\underline 0&0&\underline 0&0&\underline 0&0&\underline
 0&0&\underline 1\\ \cline{2-11}
\end{tabular}\end{center}
\begin{center}\begin{tabular}{r|r@{\small\ $\vdots$ }l|r@{\small\
 $\vdots$ }l|r@{\small\ $\vdots$ }l|r@{\small\ $\vdots$ }l|r@{\small\
 $\vdots$ }l|}\cline{2-11}
$M_t^2=$&0&\underline 0&0&\underline 1&0&\underline 0&0&\underline
 0&0&\underline 0\\ \cline{2-11}
\end{tabular}\end{center}
\begin{center}\begin{tabular}{r|r@{\small\ $\vdots$ }l|r@{\small\
 $\vdots$ }l|r@{\small\ $\vdots$ }l|r@{\small\ $\vdots$ }l|r@{\small\
 $\vdots$ }l|}\cline{2-11}
$M_u^1=$&0&\underline 1&0&\underline 0&0&\underline 0&0&\underline
 1&0&\underline 0\\ \cline{2-11}
\end{tabular}\end{center}
\begin{center}\begin{tabular}{r|r@{\small\ $\vdots$ }l|r@{\small\
 $\vdots$ }l|r@{\small\ $\vdots$ }l|r@{\small\ $\vdots$ }l|r@{\small\
 $\vdots$ }l|}\cline{2-11}
$M_e^1=$&0&\underline 0&0&\underline 0&0&\underline 1&0&\underline
 0&0&\underline 0\\ \cline{2-11}
\end{tabular}
\end{center}
where $\underline 0=0000$ and $\underline 1=1111$.

Mask $N_\sigma$ is the complement of $\sum_j M_\sigma^j$, preparing second
type computations. The operation $L\& N_\sigma$ will erase the blocks for
which $\sigma= tr[i]$. For our example, we have:

\begin{center}
\begin{tabular}{r|r@{\small\ $\vdots$ }l|r@{\small\
 $\vdots$ }l|r@{\small\ $\vdots$ }l|r@{\small\ $\vdots$ }l|r@{\small\
 $\vdots$ }l|}\cline{2-11}
$N_t=$&0&\underline 1&0&\underline 0&0&\underline 1&0&\underline 1&0
&\underline 0\\ \cline{2-11}
\end{tabular}\end{center}
\begin{center}
\begin{tabular}{r|r@{\small\ $\vdots$ }l|r@{\small\
 $\vdots$ }l|r@{\small\ $\vdots$ }l|r@{\small\ $\vdots$ }l|r@{\small\
 $\vdots$ }l|}\cline{2-11}
$N_u=$&0&\underline 0&0&\underline 1&0&\underline 1&0&\underline 0&0
&\underline 1\\ \cline{2-11}
\end{tabular}\end{center}
\begin{center}
\begin{tabular}{r|r@{\small\ $\vdots$ }l|r@{\small\
 $\vdots$ }l|r@{\small\ $\vdots$ }l|r@{\small\ $\vdots$ }l|r@{\small\
 $\vdots$ }l|}\cline{2-11}
$N_e=$&0&\underline 1&0&\underline 1&0&\underline 0&0&\underline 1&0
&\underline 1\\ \cline{2-11}
\end{tabular}\end{center}

Generally, if $k$ is table $tr$ size,
$$M_\sigma^j=\sum_{\scriptstyle tr[i]=\sigma \hbox{ and  } pr[i]=i-j\atop
1\leq i\leq k}
\Big(\big((1\sl \Omega)-1\big)\sl {\big((\Omega+1)(i-1)\big)\Big)}.$$
and $$N_\sigma=\sum_{\scriptstyle p_i\not=\sigma\atop
 1\leq i\leq k}
\Big(\big((1\sl \Omega)-1\big)\sl
 {\big((\Omega+1)(i-1)\big)\Big)}.$$
$N_\sigma$ is the complement of $\sum_j M_\sigma^j$.

Transition $l=\langle l_1,\ldots,l_k\rangle \drv \sigma l'=\langle
                 l'_1,\ldots,l'_k\rangle$
is computed by:

$$T=\sum_j \big((L\sl j({\Omega+1}))\& M_\sigma^j\big)+(L\& N_\sigma) + E_1$$
where:
\begin{center}
\begin{tabular}
{l|r@{\small\ $\vdots$ }l|r@{\small\ $\vdots$ }l|
r@{\small\ $\vdots$ }l|r@{\small\ $\vdots$ }l|r@{\small\ $\vdots$
 }l|}\cline{2-11}
$E_1=$&0&$0001$&0&$0001$&0&$0001$&0&$0001$&0&0001\\ 
\cline{2-11}
\end{tabular}\end{center}
Adding $E_1$ amounts to add 1 to each small block.

In our example, if we scan letter $t$, the transition is computed by:
$$T=\big((L\sl 2({\Omega+1}))\& M_t^2\big)+
     \big((L\sl ({\Omega+1}))\& M_t^1\big)+(L\& N_t) + E_1$$
yielding for $l=\langle 2,5,\infty,5,\infty\rangle$, encoded by:

\begin{center}\begin{tabular}{r|r@{\small\ $\vdots$ }l|r@{\small\
        $\vdots$ }l|r@{\small\ $\vdots$ }l|r@{\small\ $\vdots$
        }l|r@{\small\ $\vdots$ }l|}
\cline{2-11}
\rule[0.4cm]{0cm}{2pt}$L=$&0&$\overline 
 {15}$&0&$\overline  {5}$&0&$\overline 
 {15}$&0&$\overline  {5}$&0&$\overline 
 {2}$\\ \cline{2-11}
\end{tabular}
\end{center}

\noindent the result:

\begin{center}\begin{tabular}{r|r@{\small\ $\vdots$ }l|r@{\small\
        $\vdots$ }l|r@{\small\ $\vdots$ }l|r@{\small\ $\vdots$
        }l|r@{\small\ $\vdots$ }l|}
\cline{2-11}
\rule[0.4cm]{0cm}{2pt}$T=$&1&$\overline 
 {0}$&0&$\overline  {6}$&1&$\overline 
 {0}$&0&$\overline  {6}$&0&$\overline 
 {1}$\\ \cline{2-11}
\end{tabular}\end{center}
All the blocks contain the correct result, except for the leftmost block and
the middle block where an overflow occurred. To treat blocks where overflow
occurred it suffices of initialise again these blocks by replacing $T$ with
$L'=T-\big((T\& E_2)\sr\Omega\big)$, where:
\begin{center}
\begin{tabular}
{l|r@{\small\
 $\vdots$ }l|r@{\small\ $\vdots$ }l|r@{\small\ $\vdots$ }l|r@{\small\
 $\vdots$ }l|r@{\small\ $\vdots$ }l|}\cline{2-11}
$E_2=$&1&\underline 0&1&\underline 0&1&\underline 0&1&\underline 0&1&\underline 0
\\ \cline{2-11}
\end{tabular}
\end{center}
We find:
\begin{center}
\begin{tabular}
{l|r@{\small\
 $\vdots$ }l|r@{\small\ $\vdots$ }l|r@{\small\ $\vdots$ }l|r@{\small\
 $\vdots$ }l|r@{\small\ $\vdots$ }l|}\cline{2-11}
$T\& E_2=$&1&\underline 0&0&\underline 0&1&\underline 0&0&\underline 0&0&\underline 0
\\ \cline{2-11}
\end{tabular}
\end{center}
Hence:
\begin{center}
\begin{tabular}
{l|r@{\small\
 $\vdots$ }l|r@{\small\ $\vdots$ }l|r@{\small\ $\vdots$ }l|r@{\small\
 $\vdots$ }l|r@{\small\ $\vdots$ }l|}\cline{2-11}
$(T\& E_2)\sr\Omega=$&0&$\overline 1$&0&\underline 0&0&$\overline 1$&0&\underline 0&0&\underline 0
\\ \cline{2-11}
\end{tabular}
\end{center}
and finally:
\begin{center}\begin{tabular}{r|r@{\small\ $\vdots$ }l|r@{\small\
        $\vdots$ }l|r@{\small\ $\vdots$ }l|r@{\small\ $\vdots$
        }l|r@{\small\ $\vdots$ }l|}
\cline{2-11}
\rule[0.4cm]{0cm}{2pt}$L'=T-\big((T\& E_2)\sr\Omega\big)=$&0&$\overline 
 {15}$&0&$\overline  {6}$&0&$\overline 
 {15}$&0&$\overline  {6}$&0&$\overline 
 {1}$\\ \cline{2-11}
\end{tabular}
\end{center}

Last we define a counter $c_i$ for each pattern $m_i$, and increment it
whenever $l_{f[i]}<w+1$, which is implanted by: 
$M_i\& L < (w+1) 2^{(\Omega+1)(f[i]-1)}$, for $i=1,\ldots,k$, where 
$M_i= \big((1\sl \Omega )-1\big)\sl {\big((\Omega+1)(f[i]-1)\big)}$.

Our algorithm treats the more complex case where we demand that all episodes
appear in a same window, a case that cannot be treated by the separate
counting of the number of windows containing each episode. A simple
modification of the counting condition enables us to also count
{\em with a single scan of the text} the number of windows containing each
individual episode, in a more efficient way than if the text were to be
scanned for each episode.

\begin{thm}\label{mult.thm} There exists an on-line algorithm in time $O(nq)$
solving the parallel search of $q$ serial episodes in a size $n$ text
(assuming the episode alphabet has at most $\sqrt n/q$ letters) on MP--RAM.
\end{thm}

\proof Let $\alpha$ be the number of letters of the alphabet. As in
\cite{dfggk}, we treat in the same way all letters not occurring in the
patterns; this leads to defining two masks $M_{\it other}$ and $N_{\it other}$
common to all such letters. Let $|w|$ be the length of the binary expansion of
$w$. The algorithm consists of four steps:

\begin{menumerate}

\item compute (at most) $q\times (k+1)$ integers representing the masks
$M_\sigma^j$, $(k+1)$ integers representing the masks $N_\sigma$ and the
integers $\Omega,\Delta, I_0,F,E_1,E_2$; all these integers are of size
$k(|w|+2)$ and are computed {\em simultaneously} in $k$ iterations at most. The
integer $k$ is the size of the trie representing the patterns:
$k \leq \sum_{i=1}^q |m_i| \leq\sqrt  n$.

\item let $c=0$ ($c$ is the number of $w$-windows containing all the patterns).

\item let $L=I_0$.

\item scan text $t$; after scanning $t_i$, compute the new state $L$
({\em on-line and  without  preprocessing} with an MP--RAM) and if $c_i<w$ for
$i=1,\ldots,q$, increment $c$ by 1.
\end{menumerate} 
Our algorithm uses only the simple and fast operations $\&$, together with
a careful implementation of $\sl,\sr$ and 
addition. Step 1 of preprocessing is in time $qk(k+1)+q(k+1)+\log(w)
\leq q(\sqrt  n)^2 + 2q\sqrt  n + q + \log(w)=O(nq)$;
in general, $k$, $q$ and $w$ are smaller than $n$ by several orders of
magnitude and we will have: $qk(k+1)+q(k+1)+\log(w)=o(n)$. In step 4 we scan
text $t$ linearly in time $O(n)$ and perform $q$ comparisons (one for each
counter $c_i$). Complexity is thus in time $nq$, hence finally a time
complexity $O(nq)$ for the algorithm. \QED

\section{Experimental results}
\label{sect.exp}

The algorithm on  MP--RAM has a better  complexity  than the
standard  algorithm, however,  the underlying computation models being different,
 we  checked experimentally that the MP--RAM algorithm
is faster. We implemented all  algorithms in C++.
Experiments were realised on a PC (256 Mo, 1Ghz) with
Linux.
The text was a randomly generated file. We 
measured the  time   with machine clock ticks.

For searching multiple   patterns, we took 3 to 5 patterns of
length 2 to 4; in  figure \ref{patternmult.fig},   case (a) is the  case
of patterns having no common prefix, and  case (b) is the  case of  patterns
having  common prefixes. In  case (a),  the MP--RAM algorithm
where  we concatenate the patterns is at least  twice as fast as the standard
``naive'' algorithm where  patterns are concatenated; both standard
algorithms (with patterns concatenated or organised in a trie)
are equivalent, the algorithm with concatenation being slightly faster; this
was predictable since a trie  organisation will not give a significant advantage
in that case; the MP--RAM algorithm
where  the patterns are organised in a trie is 30 to  50\% faster than the standard
algorithm  with trie, and   10 to  15\% slower than the  MP--RAM
algorithm where  the patterns are  concatenated. However, as soon as the total
length of the patterns is larger than 7 or 8, or  the window size is
larger than 30, if  patterns are concatenated, the automaton  state
can no longer be encoded in a single   32 bits memory cell, 
and it is better to use the MP--RAM  algorithm with trie (figure
\ref{patternmult.fig} case (b)). Figure \ref{patternmult.fig} case
(b) shows that, for patterns having common prefixes, the MP--RAM algorithm
with trie is  1.3 to  1.5 times faster than the standard algorithm
with trie, itself 1.4 to  1.6 times faster than the standard algorithm
 with concatenation.

\begin{figure*}
\BoxedEPSF{patternmult.eps scaled 520}
\caption{\small The  continuous thin lines  represent  the  execution time
 of  the MP--RAM algorithm (with trie); the dotted line
represents the  execution time   of  the MP--RAM algorithm (with
concatenation); the dashed lines  the  execution time   of  the standard algorithm
(with concatenation) and  the continuous thick lines the execution  time
of the standard algorithm (with trie).} 
\label{patternmult.fig}\end{figure*}

\section{Conclusion}

We   presented new algorithms  for multiple episode search, much more efficient than the standard algorithms.
This was confirmed by our experimental analysis. Note that with
our method, counting  the  number of  windows containing several
 episodes is not harder than  checking the existence of one
window containing these  episodes. This is not true with most other problems;
usually counting problems are  much harder than the
corresponding existence problems: for example, for the  
``matching with don't cares'' problem, the  existence problem  is in linear  time
while the  counting problem is in  polynomial  time  
\cite{kr} and  in the particular case of  \cite{mby}, the  existence problem 
is in  logarithmic time  while the  counting problem  is in 
sub-linear time.


\end{document}